\documentclass[onecolumn]{openjournal}
\usepackage{graphicx}

\usepackage{soul, xcolor}

\shorttitle{Kindling the First Stars II} 
\shortauthors{Wiggins et. al.}
\graphicspath{{./}{figures/}}

\begin{document}

\title{Kindling the First Stars II: Dependence of the Predicted PISN Rate on the Pop III Initial Mass Function}

\author{Alessa Ibrahim Wiggins$^{1}$}

\author{Mia Sauda Bovill$^{2}$}

\author{Louis-Gregory Strolger$^{3}$}

\author{Massimo Stiavelli$^{3}$}

\author{Cora Bowling$^{1}$}

\affiliation{$^1$Department of Physics and Astronomy, Texas Christian University, Fort Worth, TX}
\affiliation{$^2$Department of Astronomy, University of Maryland, College Park, College Park, MD}
\affiliation{$^3$Space Telescope Science Institute, Baltimore, MD}

\begin{abstract}

Population III (Pop III) stars formed out of metal free gas in minihalos at $z>20$. While their ignition ended the Dark Ages and begin enrichment of the IGM, their mass distribution remains unconstrained. To date, no confirmed Pop III star has been observed and their direct detection is beyond the reach of the James Webb Space Telescope (JWST) without gravitational lensing. However, a subset of massive Pop III stars end their lives in pair instability supernova (PISN). With typical energies of $\sim10^{53}$~erg, PISN light curve peaks are bright enough to be detectable by JWST and the Roman Space Telescope. The fundamental question of this work is whether or not observed PISN can be used as a diagnostic of the Pop III IMF.  In this work, we use a model of the formation of the first stars to determine the dependence of PISN rates at $z~>~5$ for a range of Pop III power law IMFs ($\alpha~=~0.2~-~2.35$) and, critically, the method by which the IMF is populated. At $z~>~15$, we predict typical rates of $10^{-2}~-~10^2$ per deg$^{2}$ per year which will produce $10^{-3}~-~0.1$/year in a single NIRCam pointing and $0.003~-~30$/year in a single Roman pointing with $0.1~-~1000$ per year detected in the HLTDS.  Our work highlights that theoretical modeling of PISN rates is required if upcoming PISN studies with JWST and Roman are going to constrain the Pop III IMF.

\end{abstract}

\section{Introduction}\label{SEC.Intro}

Population III (Pop III) stars are a theoretical population of metal-free stars. It is normally assumed that they formed in dark matter minihalos with masses of $10^5~-~10^6~M_\odot$ at $z > 30$ \citep{Abeletal:2002}. Their ignition ended the Cosmic Dark Ages, and their supernovae explosions enriched the interstellar and intergalactic mediums \citep{Jaacksetal:2018}. However, to date, no confirmed Pop III star or star cluster has been observed, and without gravitational lensing, those at high-$z$ remain beyond the reach of JWST \citep{Gardneretal:2006, Bovilletal:2022}.

Due to the inefficiency of the cooling of molecular hydrogen, Pop III stars are thought to be more massive than enriched Pop II and Pop I stars. Although it is widely agreed that typical Pop III stars are more massive compared to their enriched Pop II or Pop I counterparts, the precise mass distribution of Pop III stars remains uncertain. Certain investigations propose initial mass function upper limits reaching $1000~M_\odot$ or even higher \citep{Brommetal:1999,Abeletal:2000,Ohkuboetal:2009}. Conversely, other theoretical studies propose that Population III stars could exhibit a broad range of masses, potentially extending down to a solar mass or below \citep{Stacyetal:2016,Proleetal:2022,Muhammadetal:2022,Wollenbergetal:2020,Sugimuraetal:2020,Clarketal:2011,Greifetal:2011,Susa:2013,Yoshidaetal:2006,Parketal:2021a, Parketal:2021b}. Consequently, our understanding of both the shape of the Pop III stellar initial mass function (IMF) and the maximum mass for a Pop III star remain elusive. While their formation and evolution are likely distinct from that of enriched stars in the local Universe, they remain observationally unconstrained.

Directly detecting Pop III stars or Pop III star clusters at high-$z$ is challenging as their magnitudes are too faint for even the deepest JWST  observations \citep{Rydbergetal:2013, Bovilletal:2022}. Nevertheless, utilizing gravitational lensing shows promise, and has been studied with a focus on magnification in known lensing clusters  \citep{Zackrissonetal:2015,Windhorstetal:2018}. Detecting Pop III stars through gravitational lensing necessitates accurate lensing models, which are available for various lensing clusters \citep{Lametal:2014,Diegoetal:2016,Diegoetal:2015b,Jauzacetal:2015a,Jauzacetal:2015b}. Additionally, individual stars magnified up to 10,000 times have been observed in caustics \citep{Vanzellaetal:2020,Welchetal:2022}, providing evidence that these high-redshift stars could have masses exceeding $>~50~M_{\odot}$  \citep{Welchetal:2022}. The volume probed by very high magnification lensing clusters is modest, which may make the direct observation of a Population III star unlikely.

Consequently, one of the most favorable methods to detect the earliest stars is indirectly through their supernovae. Specifically, a subgroup of Pop III stars with masses ranging from $140~-~260~M_\odot$ undergo what is known as a pair instability supernova (PISN) \citep{FowlerH:1964,Barkatetal:1967,Rakavyetal:1967}. These explosive events release energy, perhaps exceeding $10^{53}$~erg, making them more than a hundred times more energetic than a typical core collapse supernova, although see \citet{Schulzeetal:2023} and \citet{Gal-Yam:2009ef} for a highly likely candidates with lower energy. As a result, their peak luminosities are significantly high, enabling the possibility of mass detection using both the JWST and the Roman Space Telescope~\citep{Whalenetal:2013a,Wangetal:2017,Moriyaetal:2022a,Vendittietal:2024}. This is especially true for the expected depth ($\sim28-29$ mag) and coverage($5-12$ deg$^2$) of the Roman High Latitude Time Domain Survey~\citep{Roseetal:2021}.

In addition to the unconstrained probability of a Pop III star forming in the PISN range, additional uncertainties need to be incorporated into the interpretation of observed PISN rates. Thus theoretical modeling, specifically that which can explore a multidimensional parameter space, is a critical component of any PISN survey. Currently, PISN rates have been modeled for a range of Pop III IMFs \citep{LazarB:2022,Brieletal:2022,Vendittietal:2024}. 

In this work, we present predictions for the intrinsic PISN rates for various Pop III IMFs using a range of methods to populate the IMF. We also introduce a Monte Carlo method for stocastically populating the Pop III IMF to account for the limited fragmentation of primordial gas \citep{Stiavelli:2009,Bovilletal:2022}. In Section 2, we describe the simulations and the relevant components of our model and the calculation of the PISN rates. In Section~\ref{SEC.Results}, we present our predictions for the dependence of PISN rates on the Pop III IMF. Finally, we discuss our conclusions in Section~\ref{SEC.Concl}.

\section{Simulations and Models}\label{SEC.Models}

To calculate PISN rates in a cosmological volume we build a set of merger trees using the N-body simulations described in \cite{Bovilletal:2022} with a comoving boxsize of 2 Mpc/h on a side run from $z~=~150$ to $z~=~6$ with $N~=~512$ and a dark matter particle of $\sim~10^3~M_\odot$. We assume a WMAP9 cosmology ($\Omega_M~=~0.279,~ \Omega_\Lambda~=~0.721,~h_o~=~0.7$) \citep{WMAP9}. At our redshifts of interest ($z~\sim~20~-~6$), the physical size of our box is $\approx~200$~kpc on a side. We robustly resolve halos with $M~>~10^5~M_{\odot}$, including all potential sites of Pop III star formation. In addition, the positional information for each halo allows us to track external metal enrichment by supernova ejecta from nearby halos. Simulations were run with Gadget 2 \citep{Springel:2005} on initial conditions generated by MUSIC \citep{HahnA:2011} and analyzed with Amiga \citep{KnollmannK:2009,Gilletal:2004} and consistent trees \citep{Behroozietal:2013}. 

\begin{figure}[t]
\centering
\includegraphics[width=\columnwidth]{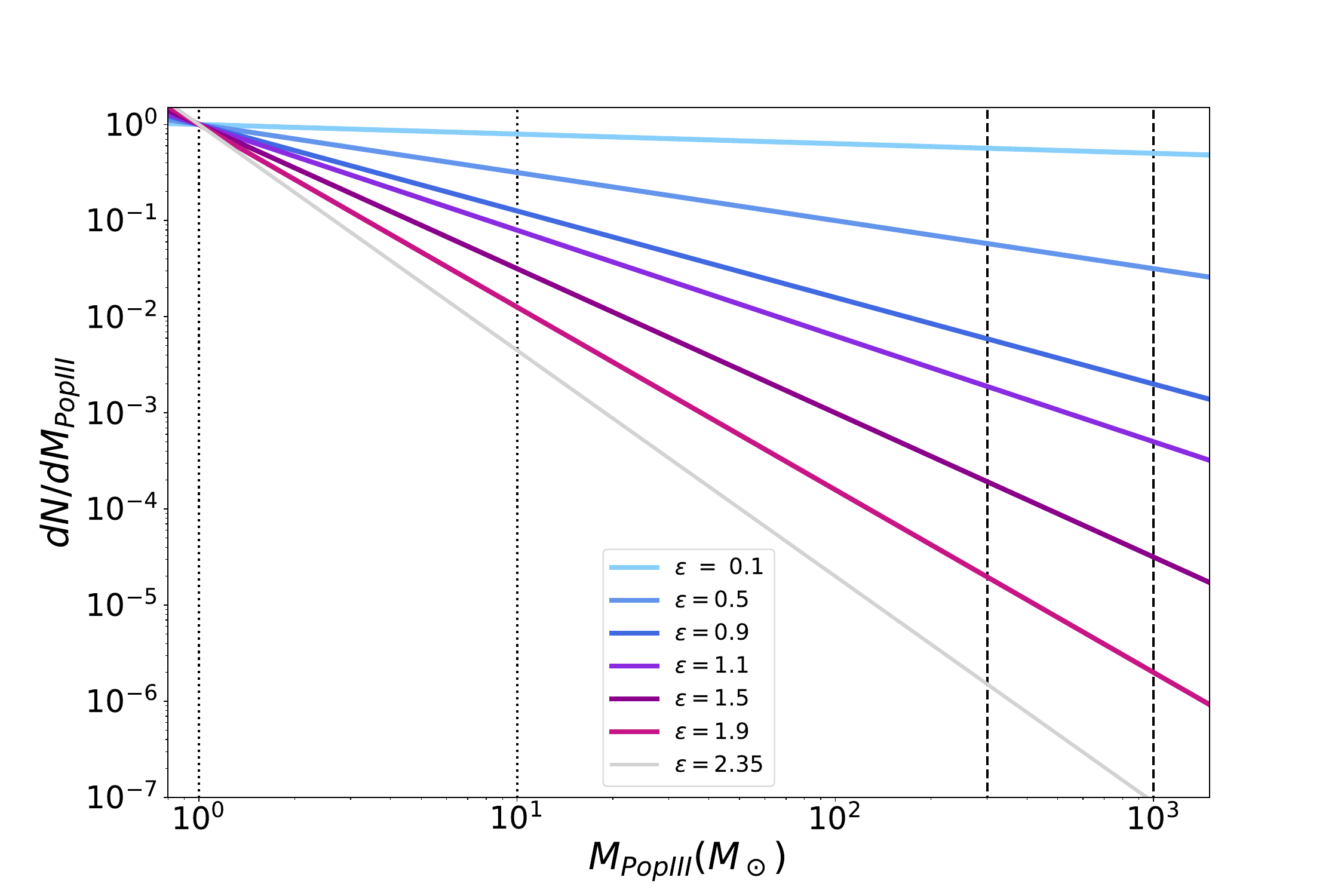}
\caption{The range of slopes for the Pop III IMF considered in this work. The color coding in this figure will be used throughout this work for the various IMF slopes. The dotted vertical lines show the minimum Pop III masses considered in this work  ($1~M_\odot$ and $10~M_\odot$) and the vertical dashed lines show the maximum Pop III masses considered in this work ($300~M_\odot$ and $1000~M_\odot$). This figure is analogous to Figure 2 in \cite{Bovilletal:2022}.}
\label{FIG.III_IMF}
\end{figure}
\vskip0.15cm

Baryonic properties are added using the model described in \cite{Bovilletal:2022} which models the Pop III star formation in a cosmological context. The model assumes Pop III stars only form in halos which are pristine, having not been enriched internally by previous episodes of star formation, nor externally by supernova ejecta from nearby halos. Only Pop III stars in the model can produce PISN. We assume the Pop III stars form with an IMF which is a power-law $dN/dM = AM^{-\alpha}$ with a given Pop III minimum mass ($m_{min}$) and maximum mass ($m_{max}$). Due to the uncertainty in the slope of the Pop III IMF, we explore a range of slopes, with $\alpha = [0.1, 0.5, 0.9, 1.1, 1.5, 1.9, 2.1, 2.35]$, as shown in Figure~\ref{FIG.III_IMF}. The most bottom heavy slope of $\alpha =2.35$ corresponds to the Salpeter IMF for Pop I stars \citep{Salpeter:1955}. The evolution of the cosmic star formation history for the redshift range considered here is automatically included in the \cite{Bovilletal:2022} model.
 
\subsection{Calculating $N_{\rm PISN}$}

Before we can determine the rate of PISN, we first need to determine the number of stars which form in the PISN mass range ($140~-~260~M_\odot$) in a specific halo for a given Pop III IMF. To do this, we populate the Pop III IMF in three different ways, through mass normalization, number normalization, and through monte carlo population. For all three, we assume that every star which forms in the PISN mass range explodes as a PISN with an energy of $10^{53}$~erg.

\subsubsection{Mass Normalization}
\label{SEC.A_M}

The first method we use to populate the Pop III IMF is the one commonly used to calculate PISN rates for various potential Pop III IMFs \citep{LazarB:2022,Brieletal:2022,Vendittietal:2024}. The number of PISN is determined by the integral of $dN/dm$ for $140~M_\odot~\le~m~\le~260~M_\odot$. In this first option the IMF is normalized by the maximum possible mass in Pop III stars, $M_{\rm III}^{max}$,
\begin{equation}
	M_{\rm III}^{max} = \epsilon_{\rm III}f_bM_{vir} = \int^{m_{max}^{\rm III}}_{m_{min}^{\rm III}} A_M m^{\alpha + 1}dm,
	\label{EQ:MIII}
\end{equation}
where $M_{vir}$ is the virial mass of the halo, $f_b$ is the cosmic baryon fraction, $\epsilon_{\rm III}$ is the star formation efficiency of Pop III stars, and $m_{max}^{\rm III}$ and $m_{min}^{\rm III}$ are the maximum and minimum possible Pop III masses respectively. We use $\epsilon_{\rm III}~=~0.01$ which is the fiducial value used in \cite{Bovilletal:2022}. For the power law IMF assumed in this work, a halo with $M_{vir}$ gives us a $N_{\rm PISN}$ of
\begin{equation}
N_{\rm PISN} = \int^{260~M_\odot}_{140~M_\odot} A_M m^{\alpha}dm,
\end{equation}
where $A_M$ is the constant for a mass normalization of the Pop III IMF.  When the Pop III IMF is populated this way two halos with the same $M_{vir}$ will always have the same $N_{\rm PISN}$. In addition, $N_{\rm PISN}$ is the sum of the probabilities of a star forming between $140~-~260~M_\odot$. Critically, this allows for a non-integer $N_{\rm PISN}$, including one which is less than one.

\subsubsection{Number Normalization}
\label{SEC.A_N}

For this method, we use the assumption from \cite{Bovilletal:2022} that inefficient cooling means primordial gas will not fragment as effectively as enriched gas, limiting the total number of fragments out of which Pop III stars can form. This places a limit on the {\it{number}} of Pop III stars which can form in a given halo.

Jeans collapse allows us to estimate the maximum number of fragments in a halo with $M_{vir}$ at a given redshift, $z$, assuming an initial gas temperature of $T_{vir}$ cooling to a final temperatures, $T_f$. If the gravitational collapse timescale, $\tau_{coll}$, is less than the cooling time scale, $\tau_{cool}$ in the halo, the gas will collapse into a single fragment. However, for Pop III forming halos with $M~\gtrsim~10^6 M_\odot ((1+z)/31)^{-2.07}$, the collapse timescale is longer than the cooling timescale. Therefore, the gas cools faster than it collapses, resulting in additional fragmentation \citep{Bovilletal:2022,Stiavelli:2009}. Note, in this work, we are not accounting for more complex physics such as increases in fragmentation with the angular momentum of the halo, however, our simple model is able to reproduce the Pop III star formation efficiencies from \cite{SkinnerW:2020} \citep{Bovilletal:2022}. 

The maximum number of fragments, $N_{frag}$ which can form from the primordial gas in a halo is given by
\begin{equation}
    N_{frag} \le 9.12 \bigg(\frac{M_{vir}}{10^6 M_{\odot}}\bigg)^{4/3} \bigg(\frac{1+z}{31}\bigg)^2  = \int^{m_{max}^{\rm III}}_{m_{min}^{\rm III}} A_N m^{\alpha}
    \label{EQ:Nfrag}
\end{equation}
where $M_{vir}$ is the virial mass of the halo and $z$ is the redshift of the halo. For the power law IMF assumed in this work, a halo with $M_{vir}$ at $z$ gives us a $N_{\rm PISN}$ of

\begin{equation}
N_{\rm PISN} = \int^{260~M_\odot}_{140~M_\odot} A_N m^{\alpha} dm
\end{equation}

where $A_N$ is the normalization constant for normalization by number. As with \S~\ref{SEC.A_M}, when the Pop III IMF is populated this way two halos with the same $M_{vir}$ at the same $z$ will always have the same $N_{\rm PISN}$. Once again, $N_{\rm PISN}$ is the sum of the probabilities of a star forming between $140~-~260~M_\odot$ allowing for a non-integer $N_{\rm PISN}$, including values which are less than one.

\subsubsection{Monte Carlo Population of the IMF}
\label{SEC:MonteCarlo}

Our final method of populating the Pop III IMF is to use a Monte Carlo approach to populate the number of stars in the IMF. This method is more realistic for several reasons. First, the specific number and masses of stars at the high mass of the IMF is dominated by Poisson noise. Specifically, for low number statistics dominated by poisson noise, a large number of small probabilities summing to provide certainty of an event may not be an accurate model for the formation of massive stars. In addition, the statistical approach to populating the IMF assumes it is possible to have a have a fraction of a PISN {\it{and}} that when you add up all of those fractions, they add constructively. 

The method we use to populate the IMF using a Monte Carlo technique is the same used in \cite{Bovilletal:2022} and we describe it below.~\begin{enumerate}

\item First, we randomly populate a given Pop III IMF with a million stars of masses, $m_{III}^i$. As in \S~\ref{SEC.A_M} and \S~\ref{SEC.A_N}, the IMF is defined by the slope of a power law and the minimum and maximum masses of the Pop III stars.

\item We then calculate the cumulative sum of the $10^6$ Pop III `stars'  where $M_{III}^j~=~\sum_{i<j} m_{III}^i$ and truncate the number of 'stars' when $M_{III}^j~<~M_III^{max}$ where $M_{III}^{max}$ is the maximum mass in Pop III stars which can be formed in a halo of a given $M_{vir}$.

\item Finally, we account for the limited fragmentation of primordial gas by truncating stars a second time so that the total number of Pop III stars, $N_{\rm III}~<~6 N_{frag}$, where $N_{frag}$ is the integer value of the maximum number of fragments in a primordial gas for a halo with a given $M_{vir}$ and $z$ (Equation~\ref{EQ:Nfrag}), and 6 is the maximum number of Pop III stars which form per fragment \citep{Susaetal:2014}.

\end{enumerate}

The number of PISN is determined by how many of the remaining stars are between $140~M_\odot$ and $260~M_\odot$. Unlike when the IMF is populated statistically, this will always be an integer number and be different each time the IMF of a halo is populated with a different initial set of `stars.'

\begin{figure*}[h]
\centering
\includegraphics[width=\columnwidth]{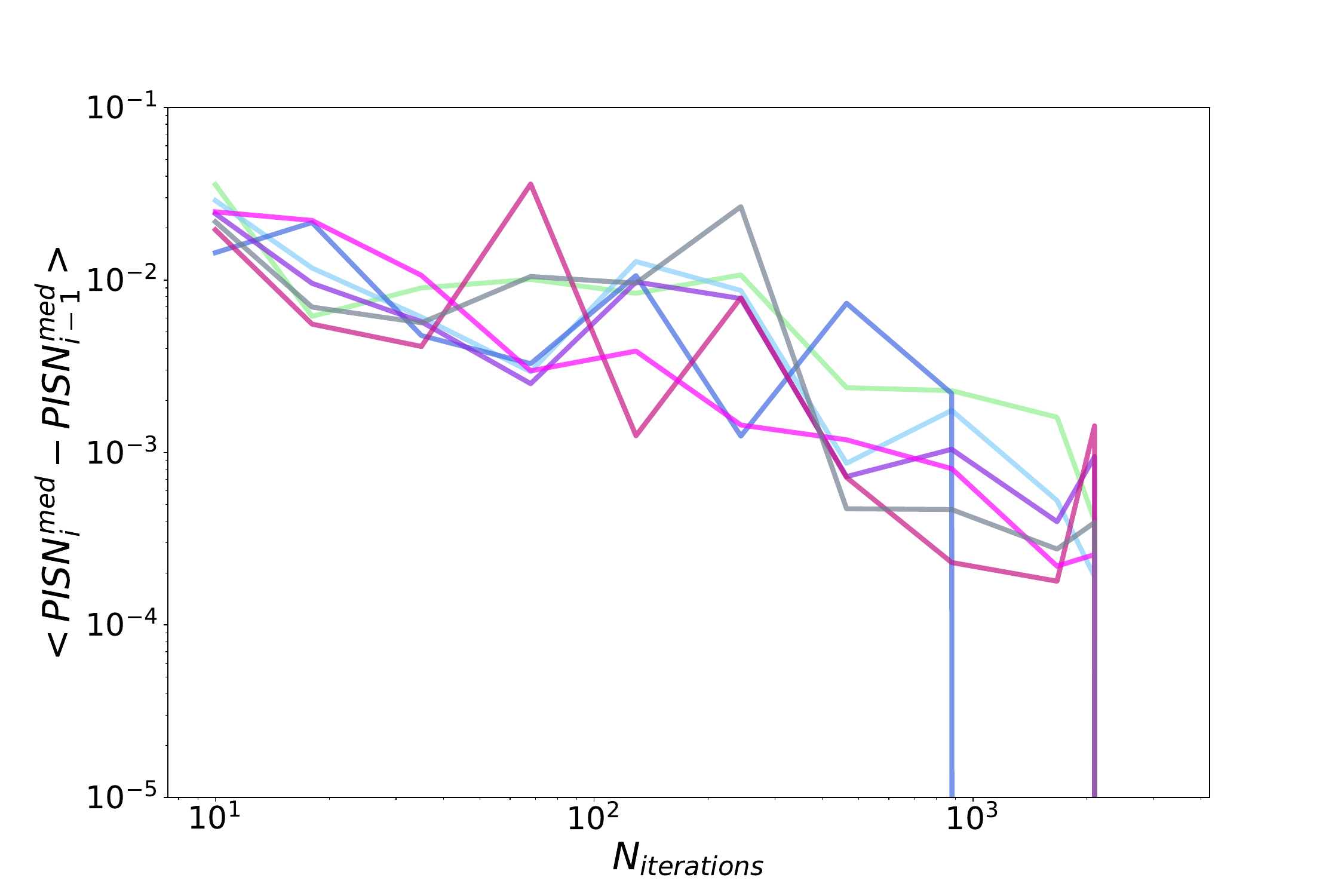}
\caption{Mean of the difference between the median of the $N_{\rm PISN}$ per halo versus, $N$, the number of times we use our Monte Carlo method to populate the IMF for $N = (10, 18, 35, 68, 129, 244, 464, 879, 1668, 3162)$. This shows that for $N~=~1000$ we are convergent to within $10^{-3}$, which is two order of magnitude below the difference in PISN rates of similar Pop III IMFs.}
\label{FIG.convergence}
\end{figure*}

We populate the IMF $N$ times and determine the median and standard deviation of $N_{\rm PISN}$ in each halo. For a specific Pop III IMF, the median $N_{\rm PISN}$ is considered the typical $N_{\rm PISN}$ for a given halo and the standard deviation measures the inherent scatter in $N_{\rm PISN}$ for a given halo. The latter is critical for the interpretation of PISN rates as it quantifies whether there is an {\it{observable}} difference in the PISN rate for various Pop III mass distributions.

We choose 1000 iteration of the Monte Carlo method because it provides convergence which is an order of magnitude less than the inherent scatter in the number of PISN for a given Pop III IMF. In Figure~\ref{FIG.convergence}, we demonstrate this by plotting the average difference in the median PISN for 10 - 3000 iterations. For N<1000 iterations the change in the median PISN rate would not produce any meaningful change in our predictions. 

\subsection{PISN Rate Calculation}

In this section, we describe how we converted the number of PISN per halo in our models into predictions for the observed PISN rate per square degrees (Figure ~\ref{FIG.PISN_RATES}).  After populating the Pop III IMF, we first determine the total number of PISN in each halo for every snapshot in the N-body simulation. Second, we smooth this in redshift using a kernel with $\Delta z~=~0.5$, where our chosen $\Delta z$ is determined by the typical band width of the NIRCam wide filters on JWST for emission in the rest frame UV \citep{Rieke:2008am}.

The next step is to calculate the number density of PISN at a given $z$ per square degree by dividing the smoothed number of PISN by the angular size of the simulation box at that redshift. The penultimate step is to transform the number density of PISN on the sky at a given redshift to a PISN {\it{rate}} per square degree. We then divide the number density of PISN at each redshift by the time between simulation outputs ($10$~Myr). Finally, we add the effects of time dilation due to the expansion of spacetime  by multiplying the PISN rates by $(1~+~z)$ \citep{WeinmannL:2005}.

\section{Results}\label{SEC.Results}

The fundamental question of this work is whether or not observed PISN can be used as a diagnostic of the Pop III IMF. The short answer to this question is yes, however there are important caveats.

Calculating a PISN rate from an underlying population of Pop III stars requires assumptions about how the high mass end of that IMF is populated. Therefore, how the PISN rates detected by JWST and Roman can be interpreted depends on what assumptions the models make. 

\begin{figure*}[h]
\centering
\includegraphics[width=\textwidth]{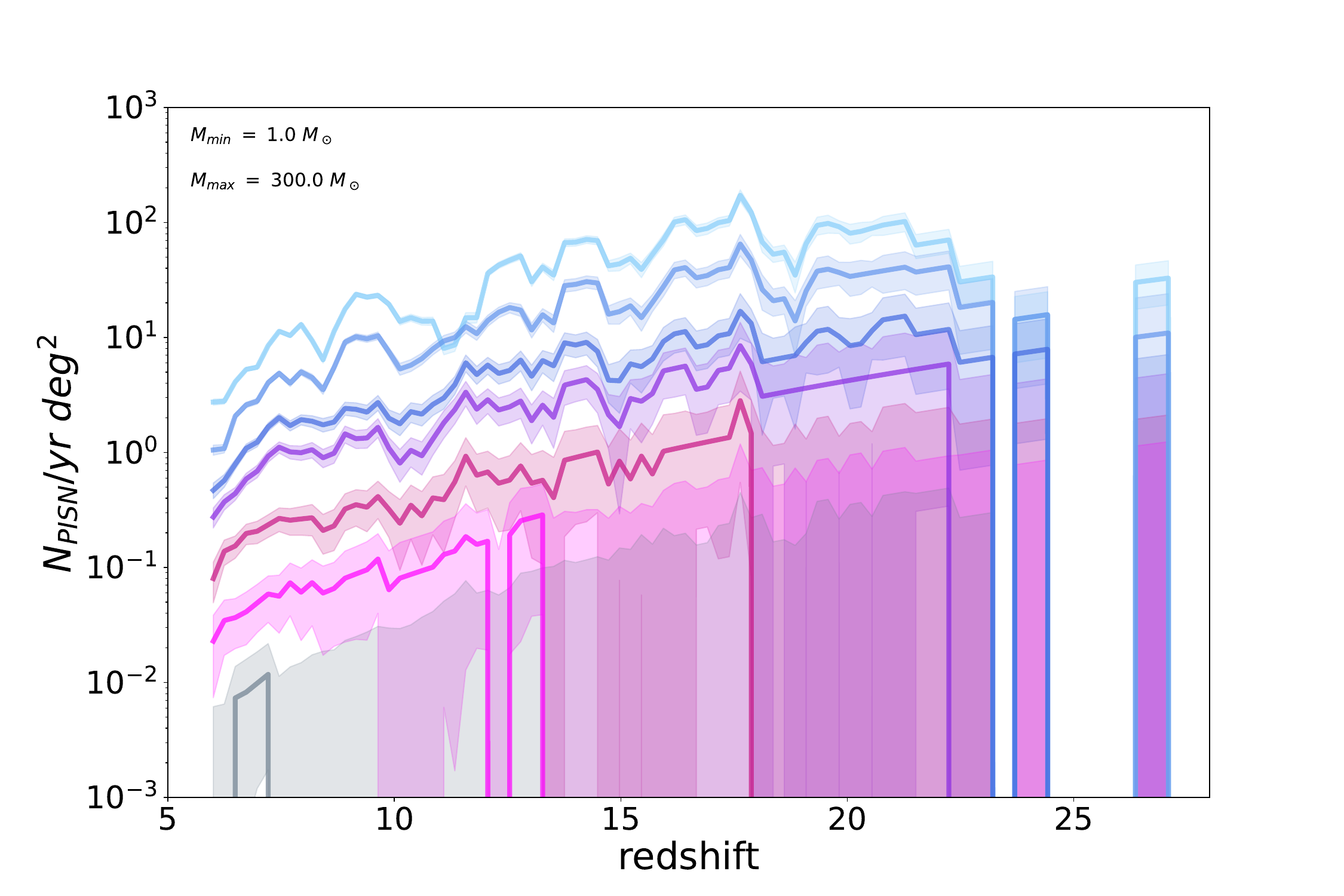}
\caption{The PISN rates calcuated using a Monte Carlo population of the Pop III IMF for $m_{min}^{\rm III}~=~1.0~M_\odot$ and $m_{max}^{\rm III}~=~300~M_\odot$. Solid lines show the median PISN rate for 1000 iterations of the IMF population, and the shaded region show the expected range of PISN rates for each IMFs. The colors correspond to the IMF slopes in Figure~\ref{FIG.III_IMF}.}
\label{FIG.PISN_RATES}
\end{figure*}

We begin with the PISN rates calculated from Pop III IMFs populated with the Monte Carlo method described in \S~\ref{SEC:MonteCarlo}. The median PISN rates in Figure~\ref{FIG.PISN_RATES} show a spread of 4 - 5 orders of magnitude for $\alpha = 2.35 - 0.1$. 

However, when the Pop III IMF is populated using our Monte Carlo approach, a halo with a given $M_{vir}$ at a given redshift, $z$, will not always have the same $N_{\rm PISN}$. This produces an inherent scatter in the PISN rate for a given Pop III IMF, quantified by the standard deviation in the PISN rate and shown as the shaded regions in Figure~\ref{FIG.PISN_RATES}. Critically, in order for observed PISN rate to be an effective tool to constrain the Pop III IMF, the difference between PISN rates for different Pop III IMFs must be {\it{greater}} than the inherent scatter in the PISN rate for each IMF. This is illustrated by the comparisons between the bottom heavy ($\alpha = (2.35,~2.1,~1.9)$ and top heavy ($\alpha = 1.5 - 0.1$) IMFs in Figure~\ref{FIG.PISN_RATES}.

For all IMFs considered in this work, the difference in the median modeled PISN rates is large enough to allow PISN rates observed by JWST and Roman to provide significant constraints. When the inherent scatter for a given IMF is taken into account the picture is more complex. 

\begin{figure*}[h]
\centering
\includegraphics[width=0.49\textwidth]{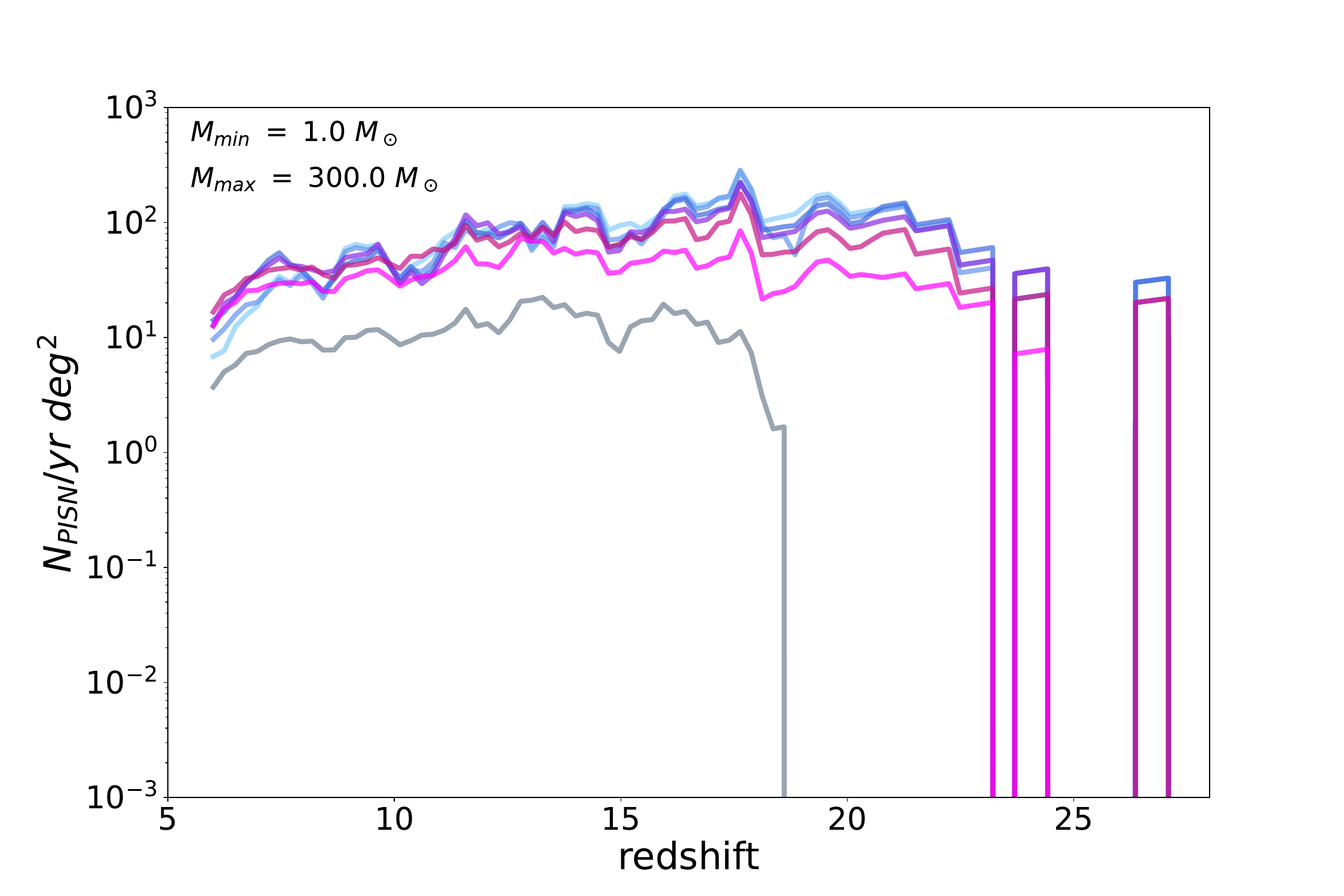}
\includegraphics[width=0.49\textwidth]{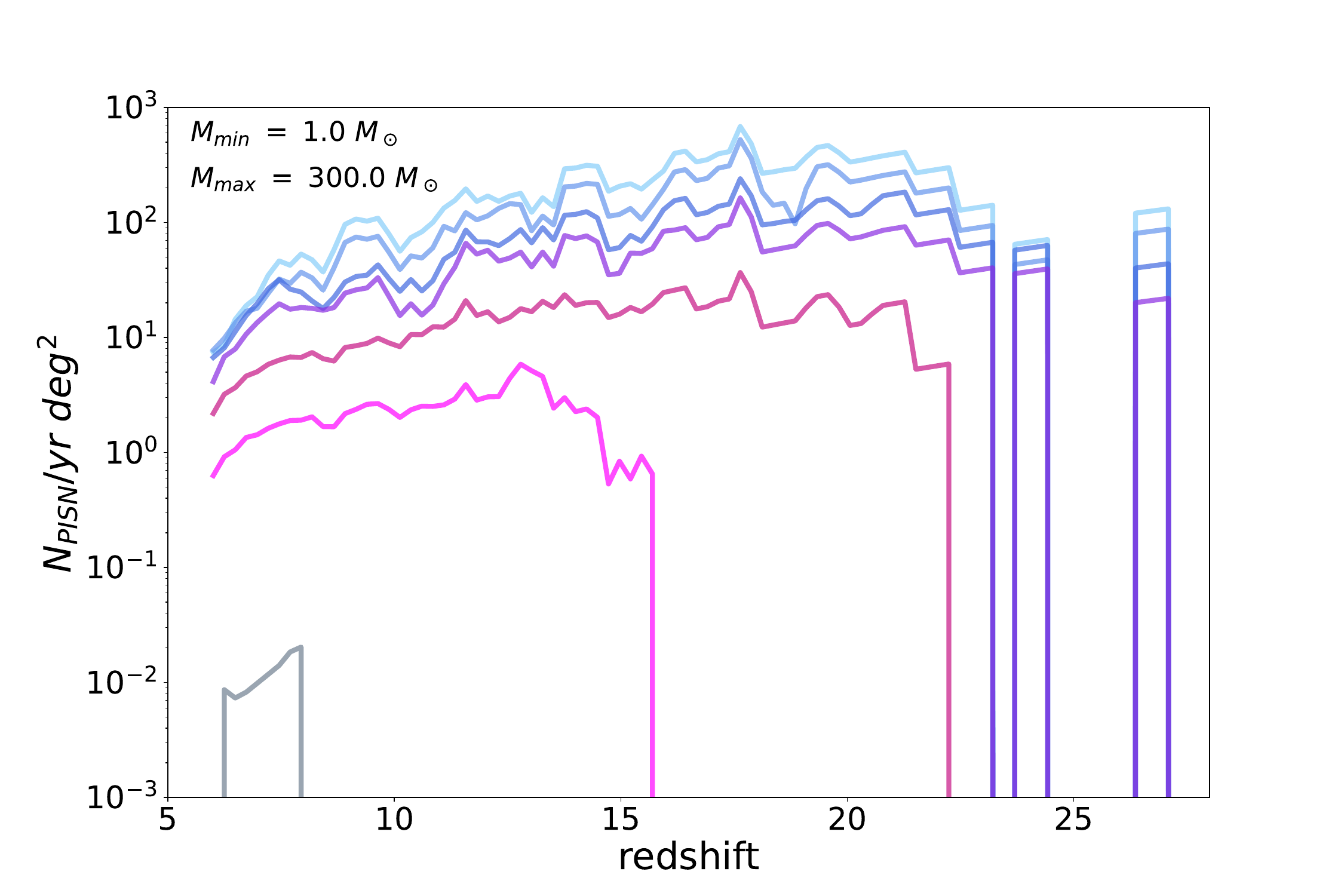}
\caption{{\it{(Left))}} The PISN rates for the Pop III initial mass functions show in Figure~\ref{FIG.III_IMF} assuming the Pop III IMF is normalized by mass. {\it{(Right))}} PISN rates for the Pop III initial mass functions show in Figure~\ref{FIG.III_IMF} assuming the Pop III IMF is normalized by number.}
\label{FIG.PISN_Mass}
\end{figure*}

For the top heavy IMFs the inherent scatter is smaller than the differences between the median PISN rates. Therefore the PISN rate can be used to differentiate between various top heavy IMFs at $z~<~15$. For bottom heavy IMFs the inherent scatter is {\it{larger}} than the differences in the PISN rates except for $z~<~9-10$. Thus, observed PISN rate will be of only limited use when differentiating between bottom heavy Pop III IMFs. In addition, at $z~>~15$, the inherent scatter is greater than the differences for all IMFs considered in this work, severely limiting the efficacy of observed PISN rates as a constraint on the Pop III IMF.

We next quantify how our Monte Carlo method of populating the IMF changes the predicted PISN rates compared to the statistical method used in other work. Figure~\ref{FIG.PISN_Mass} shows the PISN rates calculated from a statistical population of a Pop III IMF normalized by the total mass of Pop III stars and the number of Pop III stars as described in \S~\ref{SEC.A_M} and \S~\ref{SEC.A_N}.

When the IMF is statistically populated, and normalized by mass, we find PISN rates of $10^1~-~10^2$ per square degree per year with minimal dependence of the PISN rate on the Pop III IMFs for the IMFs probed in this work

\begin{figure*}[h]
\centering
\includegraphics[width=\textwidth]{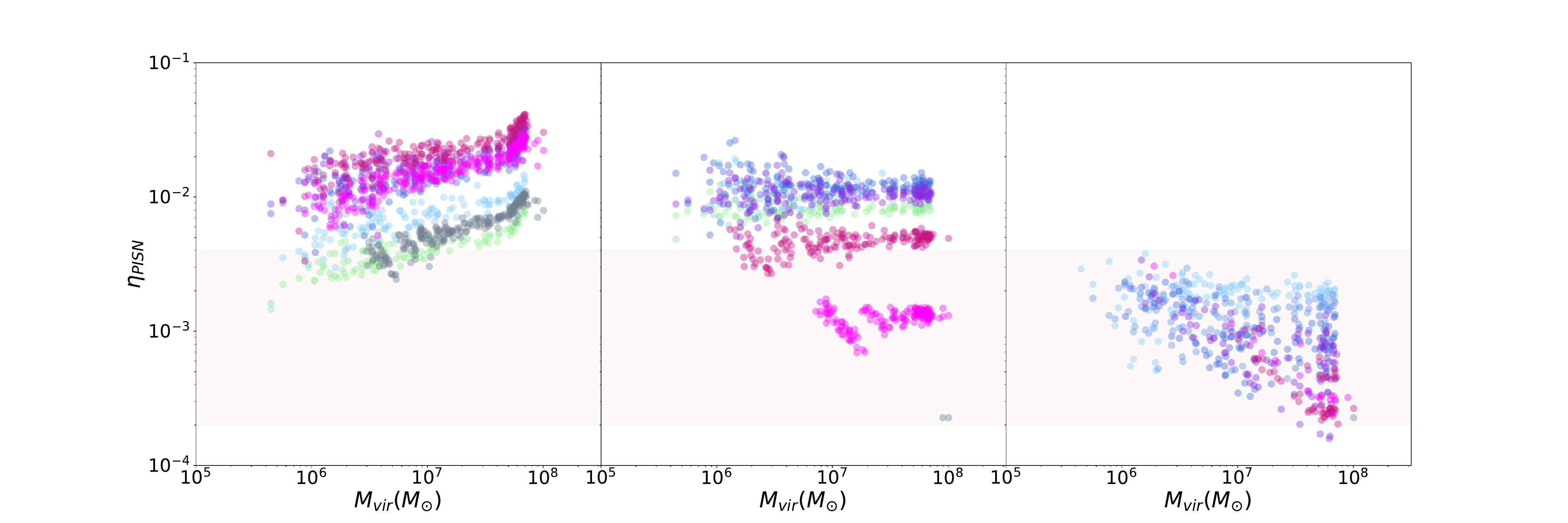}
\caption{The efficiency of PISN formation for the three methods we use to populate the Pop III IMF, normalized by mass (left), normalized by number (center), and using our Monte Carlo approach (right). The colors of the points correspond to the Pop III IMFs shown in Figure~\ref{FIG.III_IMF} and the shaded region shows the PISN efficiency range given in \cite{LazarB:2022}.}
\label{FIG.PISN_efficiency}
\end{figure*}

As seen in Figure~\ref{FIG.PISN_efficiency}, we find that our results are in broad agreement with \cite{LazarB:2022}. For our mass normalized models, our slightly higher values are likely due to the lack of an exponential cut off in our Pop III IMFs. However, for all three methods, we produce PISN efficiencies, $\epsilon_{\rm PISN} = N_{\rm PISN}/M_{\rm III}$, which are in agreement with the range reported in their work. This implies that for similar Pop III IMFs, $\epsilon_{\rm PISN}$ is roughly convergent for different methods of calculating the PISN rate.

When the IMF is statistically populated and normalized by the maximum number of Pop III stars (Equation~\ref{EQ:Nfrag}), we find a stronger dependence of the modeled PISN rates on the Pop III IMF, with a spread of $2~-~3$ orders of magnitude in the PISN rates for the IMFs probed in this work. In addition, we see a stronger evolution with redshift, which is expected given the $(1~+~z)^{4/3}$ dependence on the number of Pop III star forming fragments in primordial gas.

Note, for both normalizations, populating the Pop III IMFs statistically produces roughly an order of magnitude higher PISN rates than the Monte Carlo population of the IMF.

It is critical to remember that when we observe a PISN rate all we will know is how many PISN have gone off in  a year in a given square degree of the sky in a given redshift range. Comparisons between Figures~\ref{FIG.PISN_RATES} and~\ref{FIG.PISN_Mass} demonstrate how important it is to consider how the IMF is populated before interpreting the observational PISN rates. To illustrate this we will consider a detection of approximately 1.0 PISN per year at $z~=~10$ assuming a Pop III stellar mass range of $1.0~-~300~M_\odot$.

For the mass normalized IMFs, this PISN rate is not consistent with any of the IMFs considered in this work. In fact, it is lower than that expected for even the most bottom heavy IMFs. One potential interpretation of this result is that only a fraction of Pop III stars between $140~M_\odot$ and $260~M_\odot$ explode in a PISN~\citep{Hegeretal:2003,Hegeetalr:2010}. For Pop III IMF normalized to the fragmentation of the primordial gas, this PISN rate would suggest a bottom heavy IMF equivalent to a power law with $\alpha~=~1.9$. Finally, when the IMF is populated using a Monte Carlo approach the same PISN rate would suggest a moderately top heavy IMF with $\alpha~\approx~1.0$.

\section{Discussion and Conclusions}\label{SEC.Concl}

Roman and JWST are the first telescopes with the potential to detect PISN \citep{Whalenetal:2013a,Wangetal:2017,Moriyaetal:2022a,Vendittietal:2024}. In this work, we produced predictions for PISN rates for a range of Pop III IMFs and three methods for populating the Pop III IMF. The result presented in this work highlight the critical roll theoretical modeling must play in the extraction of the astrophysics of the underlying stellar populations from the observed PISN rates.

\begin{itemize}

\item Depending on the Pop III mass distribution and the method by which we populate the IMF, we predict $10^{-2}~-~10^2$ PISN per square degree per year. This would produce $10^{-3}~-~0.1$/year in a single NIRCam pointing and $0.003~-~30$/year in a single Roman pointing with $0.1~-~1000$ per year detected in the HLTDS. In particular, for top heavy IMFs, this almost guarantees detection of a PISN with Roman's HLTDS, assuming all Pop III stars with $140~<~m~<~260~M_\odot$ explode as a PISN.

\item When done in concert with the modeling of PISN rates for a given Pop III IMF, PISN rates observed with JWST and Roman will be able to differentiate between bottom and top heavy IMFs.

\item The strong dependence of the PISN rate on the shape of the Pop III IMF and how it is populated means the details of modeling of PISN rates must be a critical component of successfully extrapolating the underlying astrophysics of the first stars from observed PISN rates. Specifically, we find populating the Pop III IMF with a Monte Carlo approach produces PISN rates which an order of magnitude lower with greater dependence on the IMF than their statistical counterparts. Therefore, the interpretation of observed PISN rates is strongly dependent on how the Pop III IMF is populated in the models.

\end{itemize}

As previously mentioned, the recent analysis of SN2018ibb as a potential PISN \citep{Schulzeetal:2023} highlights the complexities of PISN rate predictions. With $M_{peak}\approx-21.8$, SN2018ibb would be detectable to $z\sim10$, with a duration of 50-100 days. A more complex analysis which incorporates a variable PISN duration in the context of planned PISN surveys with JWST and Roman will be the subject of future study.\\

\noindent The authors acknowledge the University of Maryland supercomputing resources (http://hpcc.umd.edu) made available for conducting the research reported in this paper.

\clearpage
\bibliographystyle{mnras}
\bibliography{references}

\end{document}